# CSCO Criterion for Entanglement and Heisenberg Uncertainty Principle

J. Y. Zeng[1], Y. A. Lei[1], S. Y. Pei[2], X. C. Zeng[3]

1 School of Physics, Peking University, Beijing, 100871, China

2 Department of Physics, Beijing Normal University, Beijing, 100875, China,

3 Department of Chemistry, University of Nebraska-Lincoln, Lincoln, NE 68588, USA

Abstract

We show that quantum entanglement and the Heisenberg uncertainty principle are inextricably connected. Toward this end, a complete set of commuting observables (CSCO) criterion for the entanglement is developed. Assuming $(A_1, A_2, …)$ and $(B_1, B_2, …)$ being two CSCO's for a given system, and $C$ being the matrix, $C_{ij} = \mathrm{i}[B_i, A_j]$, for each given row $i$ ($i$=1, 2,…) if at least one matrix element $C_{ij}$ ($j$=1, 2,…) is nonzero, then for the simultaneous eigenstates $\{|\psi\rangle = |A'_1, A'_2, …\rangle\}$ of $(A_1, A_2, …)$, the simultaneous measurements of $(B_1, B_2, …)$ are, in general, entangled. The only exception is when all the simultaneous eigenstates $\{|\psi\rangle\} = \{|A'_1, A'_2, …\rangle\}$, $\langle\psi|C|\psi\rangle = 0$. This CSCO criterion may be considered as an extension of the Heisenberg uncertainty principle to quantum systems with either two (or more) particles or multi-degrees of freedom (MDF).

The uncertainty (indeterminacy) principle was uncovered by W. Heisenberg in 1927[1], which is one of the greatest milestones in the history of science. Seven years later, the term "entanglement" was coined by E. Schrödinger [2]. Earlier in the same year, the entangled states of two-particle (spin-less) systems were utilized to demonstrate the nonlocal correlation by Einstein, Podolsky, and Rosen (EPR) [3], and they showed how the strange nonlocal correlation contradicts to common sense of observation. About 30 years later, Bell proved that Einstein's point of view (the local realism) leads to algebraic predictions [4, 5] which contradict the quantum prediction. Many experiments have since been done, all consistent with quantum mechanics but not with the local realism [6-8]. Now it has been recognized that both uncertainty and entanglement are the fundamental features of quantum mechanics, which mark the strikingly different features from the classical mechanics. However, for a long time the inextricable connection between the entanglement and the uncertainty principle has not attracted much attention [9]. Recently it has been noted [10] that the uncertainty principle determines the nonlocality of quantum mechanics; i.e. quantum mechanics cannot be more nonlocal with measurements that respect the uncertainty principle. In this letter a CSCO criterion for entanglement is presented, which may be viewed as an extension of the uncertainty principle.

First, we briefly illustrate the precise meaning of entanglement. One point of view is that, contrary to wave-particle duality which is a one-particle quantum feature, entanglement requires involvement of at least two particles [11]. An important consequence of the entanglement is that for a two-particle entangled state the measurement of the state of one particle defines the state of the second particle instantaneously, whereas neither particle possesses its own well-defined state prior to the measurement [12]. Another point of view is that entanglement may apply to a set of particles, or to two or more commuting observables of a single particle [13-14]. As emphasized by V. Vedral [15], entanglement takes (at least) two to tangle, although these two need not to be particles. To study entanglement, two or more subsystems need to be identified, together with the appropriate degrees of freedom that might be entangled, where the possibly entangled degrees of freedom are called observables. A entangled state of two commuting observables $A$ and $B$ exhibits the following features [9, 12]: (a) It is not a simultaneous eigenstate of $A$ and $B$; i.e. neither observable has definite value prior to measurement. (b) The outcomes of simultaneous measurement of $A$ and $B$ are correlated with each other (in probabilistic fashion) .For example, a mesoscopic Schrödinger cat's state of a $^9Be^+$ ion was prepared[16], in which the internal (electronic) excitation is entangled with the central of mass motion. In the which-way experiment of C.Dürr et al. [17] the internal degree of freedom is entangled with the way taken by the path. The entanglement between the polarization and the linear momentum of a single photon [18], and the polarization and the angular momentum of a single photon [19] have been demonstrated experimentally. The path-spin entangled state of a single spin-1/2 particle was investigated in [20]. We prefer the later point of view for entanglement, according to which the entanglement and quantum non-locality are different resources[21].

Based on the later point of view of entanglement, we show that there exists an intrinsic relation between entanglement and the uncertainty principle. It may be noted that both entanglement and the uncertainty principle address the relation between measurements of observables. The difference is that the Heisenberg's uncertainty principle mainly stresses that two *non-commuting* observables in quantum mechanics, in general, cannot be simultaneously measured precisely, whereas the entanglement addresses the non-classical correlations between the simultaneous measurements of two or more *commuting* observables.

The uncertainty relation is usually expressed as [22-24]

$$\Delta A \Delta B \geq \frac{1}{2}|\langle C \rangle| \qquad (1)$$

Where $\Delta X = \sqrt{\langle \psi|X^2|\psi\rangle - \langle \psi|X|\psi\rangle^2}$ is the standard root-mean-square deviation of the observable $X(X = A, B)$ for a given state $|\psi\rangle$ and $\langle C \rangle = \langle \psi|C|\psi \rangle$, C is the commutator $C = i[B, A]$. The uncertainty relation (1) refers not to the precision and disturbance of a specific measurement, but to uncertainties *intrinsic* in the quantum state $|\psi\rangle$ [25], and has been experimentally verified in many settings [26]. A well-known example is that, for $A=x$, $B=p_x$ of a single particle, $C=\hbar$ (a non-zero universal constant), so for *any* quantum state of a particle the coordinate $x$ and momentum $p_x$ cannot be simultaneously measured with certainty, which is the implication of Heisenberg's uncertainty principle.

Note that the uncertainty principle *does not apparently concern with the degrees of freedom* of a given system. If two observables $A$ and $B$ are of different degrees of freedom, then $C=0$, so that $A$ and $B$ can be always measured simultaneously. In this case, there is no uncertainty regarding the measurement outcomes of $A$ and $B$. Next,we consider systems with either multi-degrees of freedom (MDF) or with many particles (MPs). It is well known that the quantum state of a MDF (or MP) system can be characterized by a simultaneous eigenstate of a CSCO [27]. The eigenstates of a CSCO span a base of representation of the Hilbert space for a given quantum system, and any state of this quantum system can be expressed as a coherent superposition of this set of simultaneous eigenstates.

We assume $(A_1, A_2, ...)$ constitute a CSCO for a given MDF system, whose simultaneous eigenstates are denoted by $\{|A_1', A_2', ...\rangle\}$, and $(B_1, B_2, ...)$ constitute another CSCO, whose simultaneous eigenstates are denoted by $\{|B_1', B_2', ...\rangle\}$. A commutator matrix $C$ is defined as

$$C_{ij} = i[B_i, A_j], \quad i, j = 1, 2, ... \quad (2)$$

Similar to the Heisenberg's uncertainty relation (1), we have

$$\Delta A_i \cdot \Delta B_j \geq \frac{1}{2}|\langle C_{ij}\rangle| \quad (3)$$

If $C_{ij} \neq 0$, the observables $A_i$ and $B_j$, in general, cannot be measured simultaneously for the state $|\psi\rangle$.

Now we present a CSCO criterion for entanglement as follows:

Assume both following conditions (a) and (b) are satisfied, i.e.

(a) for each given row $i$ ($i=1, 2, ...$) at least one of the matrix-element $C_{ij}$ ($j=1,2, ...$) is nonzero, i.e. $C_{ij} \neq 0$,

(b) for all states $\{|\psi\rangle = |A_1', A_2', ...\rangle\}$, $\langle\psi|C|\psi\rangle \neq 0$,

the simultaneous measurements of the observables $(B_1, B_2, ...)$ are correlated with each other (in probabilistic fashion); i.e. $\{|A_1', A_2', ...\rangle\}$ are entangled states of the commuting observables $(B_1, B_2, ...)$.

P roof

First, under the condition (a), $|A_1', A_2', ...\rangle$ cannot be a simultaneous state of the commuting observables $(B_1, B_2, ...)$. Next, because $(B_1, B_2, ...)$ constitute a CSCO for the given system, $|A_1', A_2', ...\rangle$ can be expressed as a coherent superposition of $\{|B_1', B_2', ...\rangle\}$

$$|A_1', A_2', ...\rangle = \sum_{B_1', B_2', ...} |B_1', B_2', ...\rangle \langle B_1', B_2', ... | A_1', A_2', ...\rangle \quad (4)$$

Under both conditions (a) and (b), all the expansion coefficients $\langle B_1' B_2' ... | A_1' A_2' ...\rangle$ are well defined and not all of them are zero. For a given state $|A_1', A_2', ...\rangle$, $\langle B_1' B_2' ... | A_1' A_2' ...\rangle$ depends on $|B_1', B_2', ...\rangle$, and $|\langle B_1' B_2' ... | A_1' A_2' ...\rangle|^2$ is just the probability of the simultaneous measurement outcomes $(B_1', B_2', ...)$ of the observables $(B_1, B_2, ...)$; i.e., $|A_1', A_2', ...\rangle$ are entangled state of the commuting observables $(B_1, B_2, ...)$.

If only the condition (a) is met, but not condition (b), i.e. for all the states $\{|\psi\rangle = |A_1', A_2', ...\rangle\}$, $\langle\psi|C|\psi\rangle = 0$, we cannot judge whether all these states, $\{|\psi\rangle = |A_1', A_2', ...\rangle\}$, are entangled, or not.

We now utilize several examples to illustrate the application of the above CSCO criterion for entanglement.

1. The EPR entangled state of a 2-particle system.

The following state of a 2-particle (spin-less) of the EPR paradox [3] is shown [9] to be a simultaneous eigenstate of the CSCO $(x, P), |x = a, P = 0\rangle$

$$\delta(x_1 - x_2 - a) = \frac{1}{\sqrt{2\pi\hbar}} \int_{-\infty}^{+\infty} dp \exp[ip(x_1 - x_2 - a)/\hbar] \quad (5)$$

Let $(A_1, A_2) = (x, P), (B_1, B_2) = (p_1, p_2)$ and $(x_1, x_2)$, we obtain, respectively,
$$C = \hbar \begin{pmatrix} 1 & 0 \\ -1 & 0 \end{pmatrix}, \quad \hbar \begin{pmatrix} 0 & 1 \\ 0 & 1 \end{pmatrix}. \tag{6}$$

Obviously, both condition (a) and (b) are obviously satisfied, so the simultaneous eigenstate (5) is an entangled state of the observables $(p_1, p_2)$ and $(x_1, x_2)$.

2. The eigenstate of total angular-momentum of a spin $1/2$ particle.

The total angular-momentum is denoted by $\boldsymbol{j} = \boldsymbol{l} + \boldsymbol{s}$, where $\boldsymbol{l}$ is the orbital angular momentum and $\boldsymbol{s}$ is the spin. The eigenstates of the CSCO $(\boldsymbol{l}^2, \boldsymbol{j}^2, j_z)$ for a given $l$ is ($\hbar = 1$)

$$|l, j, m_j\rangle = \left\{ \sqrt{\frac{l+m+1}{2l+1}} |l, m\rangle|\uparrow\rangle + \sqrt{\frac{l-m}{2l+1}} |l, m+1\rangle|\downarrow\rangle \right\}$$

$$j = l + 1/2, m_j = m + 1/2, l = 0, 1, 2, \ldots$$

$$|l, j, m_j\rangle = \left\{ -\sqrt{\frac{l-m}{2l+1}} |l, m\rangle|\uparrow\rangle + \sqrt{\frac{l+m+1}{2l+1}} |l, m+1\rangle|\downarrow\rangle \right\}$$

$$j = l - 1/2, m_j = m + 1/2, l = 1, 2, \ldots$$

(7)

For $(A_1, A_2) = (\boldsymbol{j}^2, j_z), (B_1, B_2) = (l_z, s_z)$, we have

$$C = 2\hbar \begin{pmatrix} (s_y l_x - s_x l_y) & 0 \\ (s_x l_y - s_y l_x) & 0 \end{pmatrix} \tag{8}$$

It can be shown that both condition (a) and (b) are satisfied, so all the simultaneous eigenstates (7) of $(\boldsymbol{j}^2, j_z)$ are entangled states of $l_z$ and $s_z$. In fact, the measurement outcomes of $l_z$ and $s_z$ are correlated with each other with the relative probabilities

$$\frac{(l+m+1)}{(l-m)}, \text{ for } j = l + \frac{1}{2}, \text{ and } \frac{(l-m)}{(l+m=1)}, \text{ for } j = l + \frac{1}{2}$$

3. Bell bases of a 2-qubit system.

It can be shown that for a 2-qubit system

$$\left(\sigma_x^{(1)} \sigma_x^{(2)}\right)\left(\sigma_y^{(1)} \sigma_y^{(2)}\right)\left(\sigma_z^{(1)} \sigma_z^{(2)}\right) = -1 \tag{9}$$

Hence any two of the three two-body spin-operators $(\sigma_x^{(1)} \sigma_x^{(2)})$, $(\sigma_y^{(1)} \sigma_y^{(2)})$, and $(\sigma_z^{(1)} \sigma_z^{(2)})$ constitute a CSCO. For example, let $(A_1, A_2) = \left(\sigma_x^{(1)} \sigma_x^{(2)}, \sigma_y^{(1)} \sigma_y^{(2)}\right)$, the corresponding eigenstates $|A_1', A_2', \ldots\rangle$ are just the Bell's basis

$$|\psi\rangle_{12} = \frac{1}{\sqrt{2}}[\,|\uparrow\rangle_1|\uparrow\rangle_2 \pm |\downarrow\rangle_1|\downarrow\rangle_2\,], \quad \frac{1}{\sqrt{2}}[\,|\uparrow\rangle_1|\downarrow\rangle_2 \pm |\downarrow\rangle_1|\uparrow\rangle_2\,] \tag{10}$$

For $(B_1, B_2) = (\sigma_x^{(1)}, \sigma_x^{(2)})$, $(\sigma_y^{(1)}, \sigma_y^{(2)})$, and $(\sigma_z^{(1)}, \sigma_z^{(2)})$, the corresponding $C$-matrixes are, respectively,

$$C=2\begin{pmatrix} 0 & \sigma_z^{(1)}\sigma_y^{(2)} \\ 0 & \sigma_y^{(1)}\sigma_z^{(2)} \end{pmatrix}, 2\begin{pmatrix} \sigma_z^{(1)}\sigma_x^{(2)} & 0 \\ \sigma_x^{(1)}\sigma_z^{(2)} & 0 \end{pmatrix}, 2\begin{pmatrix} -\sigma_y^{(1)}\sigma_x^{(2)} & \sigma_x^{(1)}\sigma_y^{(2)} \\ -\sigma_x^{(1)}\sigma_y^{(2)} & \sigma_y^{(1)}\sigma_x^{(2)} \end{pmatrix} \quad (11)$$

It can be shown that both condition (a) and (b) are satisfied, so the Bell bases are entangled states of $(\sigma_x^{(1)}, \sigma_x^{(2)}), (\sigma_y^{(1)}, \sigma_y^{(2)})$, and $(\sigma_z^{(1)}, \sigma_z^{(2)})$, as expected.

4. Simultaneous eigenstates of $(S^2, S_z)$ of a 2-electron system.

In the angular momentum coupling scheme, $(S^2, S_z)$ is usually adopted as a CSCO of the spin state of a 2-electron system, where $S = s^{(1)} + s^{(2)}$ is the total spin. The simultaneous eigenstates of $(S^2, S_z)$ are denoted by $|S, M_S\rangle$, $S$=1, 0, and $|M_S| \leq S$.

$$|0,0\rangle = \frac{1}{\sqrt{2}}[|\uparrow\downarrow\rangle - |\downarrow\uparrow\rangle], |1,0\rangle = \frac{1}{\sqrt{2}}[|\uparrow\downarrow\rangle + |\downarrow\uparrow\rangle]$$

$$|1,1\rangle=|\uparrow\uparrow\rangle, |1,-1\rangle=|\downarrow\downarrow\rangle \quad (12)$$

Obviously, $|0,0\rangle$ and $|1,0\rangle$ are entangled states, but $|1,1\rangle$ and $|1,-1\rangle$ are not. This can be verified by the above CSCO criterion of entanglement. For $(A_1, A_2)=(S^2, S_z)$, and $(B_1, B_2)= (s_z^{(1)}, s_z^{(2)})$, we have

$$C = -2\begin{pmatrix} s_y^{(1)}s_x^{(2)} - s_x^{(1)}s_y^{(2)} & 0 \\ -s_x^{(1)}s_y^{(2)} + s_y^{(1)}s_x^{(2)} & 0 \end{pmatrix} \quad (13)$$

It can be shown that for all the states (12), $\langle\psi|C|\psi\rangle = 0$, i.e. the condition (b) is *not* satisfied, so one cannot determine that all the states (12) are entangled, or not.

5. GHZ states

For a 3-qubit system [28]

$$\left(\sigma_x^{(1)}\sigma_y^{(2)}\sigma_y^{(3)}\right)\left(\sigma_y^{(1)}\sigma_x^{(2)}\sigma_y^{(3)}\right)\left(\sigma_y^{(1)}\sigma_y^{(2)}\sigma_x^{(3)}\right)\left(\sigma_x^{(1)}\sigma_x^{(2)}\sigma_x^{(3)}\right) = -1, \quad (14)$$

We can choose any three members of the following four 3-body spin operators

$$\{\left(\sigma_x^{(1)}\sigma_y^{(2)}\sigma_y^{(3)}\right), \left(\sigma_y^{(1)}\sigma_x^{(2)}\sigma_y^{(3)}\right), \left(\sigma_y^{(1)}\sigma_y^{(2)}\sigma_x^{(3)}\right), \left(\sigma_x^{(1)}\sigma_x^{(2)}\sigma_x^{(3)}\right)\}$$

as a CSCO. Assume

$$\{A_1, A_2, A_3\} = \{\left(\sigma_x^{(1)}\sigma_y^{(2)}\sigma_y^{(3)}\right), \left(\sigma_y^{(1)}\sigma_x^{(2)}\sigma_y^{(3)}\right), \left(\sigma_y^{(1)}\sigma_y^{(2)}\sigma_x^{(3)}\right)\}$$

$(B_1, B_2, B_3) = (\sigma_x^{(1)}, \sigma_y^{(2)}, \sigma_y^{(3)}), (\sigma_y^{(1)}, \sigma_x^{(2)}, \sigma_y^{(3)}), (\sigma_y^{(1)}, \sigma_y^{(2)}, \sigma_x^{(3)})$, we obtain the $C$-matrices, respectively,

$$2\begin{pmatrix} 0 & -\sigma_z^{(1)}\sigma_x^{(2)}\sigma_y^{(3)} & -\sigma_z^{(1)}\sigma_y^{(2)}\sigma_x^{(3)} \\ 0 & +\sigma_y^{(1)}\sigma_z^{(2)}\sigma_y^{(3)} & 0 \\ 0 & 0 & +\sigma_y^{(1)}\sigma_y^{(2)}\sigma_z^{(3)} \end{pmatrix}$$

$$2\begin{pmatrix} +\sigma_z^{(1)}\sigma_y^{(2)}\sigma_y^{(3)} & 0 & 0 \\ -\sigma_x^{(1)}\sigma_z^{(2)}\sigma_y^{(3)} & 0 & -\sigma_y^{(1)}\sigma_z^{(2)}\sigma_x^{(3)} \\ 0 & 0 & +\sigma_y^{(1)}\sigma_y^{(2)}\sigma_z^{(3)} \end{pmatrix}$$

$$2\begin{pmatrix} +\sigma_x^{(1)}\sigma_y^{(2)}\sigma_y^{(3)} & 0 & 0 \\ 0 & +\sigma_y^{(1)}\sigma_z^{(2)}\sigma_y^{(3)} & 0 \\ -\sigma_x^{(1)}\sigma_y^{(2)}\sigma_z^{(3)} & -\sigma_y^{(1)}\sigma_x^{(2)}\sigma_z^{(3)} & 0 \end{pmatrix}$$

(15)

It can be verified that both condition (a) and (b) are satisfied, so the GHZ states of 3-qubit system are entangled, as expected.

In summary, uncertainty and entanglement are two fundamental concepts in quantum mechanics. We show that there exists extricable connection between the uncertainty principle and entanglement. The uncertainty principle mainly elucidates the correlations between the simultaneous measurements of two *non-commuting* observables, whereas the entanglement describes the correlations between the simultaneous measurements of two or more *commuting* observables of a CSCO for a given system. In this work a CSCO criterion of entanglement is proposed, which may be viewed as an extension of the uncertainty principle to a MP (many-particle) or MDF (multi-degree of freedom) system.